# BRDS: An FPGA-based LSTM Accelerator with Row-Balanced Dual-Ratio Sparsification

Seyed Abolfazl Ghasemzadeh, Erfan Bank Tavakoli, Mehdi Kamal, Ali Afzali-Kusha, and Massoud Pedram

**Abstract**— In this paper, first, a hardware-friendly pruning algorithm for reducing energy consumption and improving the speed of Long Short-Term Memory (LSTM) neural network accelerators is presented. Next, an FPGA-based platform for efficient execution of the pruned networks based on the proposed algorithm is introduced. By considering the sensitivity of two weight matrices of the LSTM models in pruning, different sparsity ratios (*i.e.*, dual-ratio sparsity) are applied to these weight matrices. To reduce memory accesses, a row-wise sparsity pattern is adopted. The proposed hardware architecture makes use of computation overlapping and pipelining to achieve low-power and high-speed. The effectiveness of the proposed pruning algorithm and accelerator is assessed under some benchmarks for natural language processing, binary sentiment classification, and speech recognition. Results show that, *e.g.*, compared to a recently published work in this field, the proposed accelerator could provide up to 272% higher effective GOPS/W and the perplexity error is reduced by up to 1.4% for the PTB dataset.

## 1 INTRODUCTION

FOR applications that require processing time-dependent sequences of data such as speech recognition [1] and natural language processing (NLP) [2], Recurrent Neural Networks (RNNs) and, more specifically, Long Short-Term Memory (LSTM) Neural Networks [3] have been introduced. These networks, which create high computational loads, have to cope with resource limitation when they are implemented on hardware platforms such as FPGAs [4].

A limited number of computational resources (*e.g.*, number and type of available DSP blocks) and memory resources (*e.g.*, size and access speed of embedded block memories) in FPGAs make LSTM networks implementation challenging. To overcome this problem, several works which invoke resource sharing along with timing optimization have been suggested in the literature. Examples of these research efforts include balancing the memory bandwidth and the internal storage utilization [5], optimizing computational performance and communication requirements [6], overlapping LSTM computations with memory accesses [4], and overlapping internal computations of the LSTM architecture [7].

The weight pruning technique reduces storage and computational costs by eliminating redundant elements in the weight matrices. Reducing the model size, however, does not necessarily lead to a more efficient hardware implementation. This is because fetching unstructured pruned data may require high memory bandwidth due to the random accesses to the memory. This implies that unstructured pruning may limit the energy and performance gains that are achievable by model reduction and pruning [8]. Accordingly, when reducing the model size, the designer should try to minimize the accuracy loss due to the pruning while providing a sparsity pattern compatible with the hardware architecture.

Lower improvements achieved by unstructured pruning is due to the fact that while the input weight matrices are sparse, the input vector is dense causing rather low improvement in matrix-vector multiplication (MxV). Since some elements in a row of sparse matrix are zeros, their multiplication by their corresponding elements of the dense vector are zeros, thus not contributing to the sum in final. To avoid the useless computations, we need to access the nonzero elements requiring irregular (random) accesses to memory which is translated to inefficient utilization of the memory bandwidth and processing elements (PEs) of the NN accelerator. The reason is that the irregularity in the positions of non-zero elements in the weight matrices makes variation in the number of PEs needed for each row of the matrix inevitable, reducing the efficiency of using the process elements.

In this work, first, a Balanced Row Dual-ratio Sparsity-inducing pruning algorithm (called BRDS) is presented. In this algorithm, the input and recurrent weight matrices of the LSTM are pruned with different sparsity ratios resulting in lower accuracy loss while providing the opportunity for more weight pruning for a given target accuracy level. These two sets of matrices have different sensitivities to the pruning owing to their different contributions to the final results of the LSTM model. To lower the required memory bandwidth, the pruning is performed in a row-wise manner. Moreover, since the number of non-zero elements in each row of the sparse matrices is known at design time, number of PEs required to process the row can be determined. Both of these features provide an efficient

- *S. A. Ghasemzadeh, M. Kamal, and A. Afzali-Kusha are with School of Electrical and Computer Engineering, University of Tehran, Tehran 14399-57131, Iran. E-mail: a.ghasemzadeh@ut.ac.ir; mehdikamal@ut.ac.ir; afzali@ut.ac.ir.*
- *E. Bank Tavakoli is with the School of Computing, Informatics, and Decision Systems Engineering, Arizona State University, Tempe, AZ 85281, USA. E-mail: ebanktav@asu.edu.*
- *M. Pedram is with Department of Electrical and Computer Engineering, University of Southern California, Los Angeles, CA 90089, USA. E-mail: pedram@usc.edu.*

hardware implementation of sparse matrix-vector multiplication (SpMxV) while creating regular memory accesses for performing the operation. Next, we describe the BRDS accelerator which is an FPGA-based, row-balanced, dual-ratio sparsity-aware, low-power and high-performance architecture for the LSTM networks. The accelerator takes full advantage of the efficiency of the proposed pruning algorithm. In this accelerator, to minimize the overhead of storing the positions of non-zero values in the rows of sparse matrices, a relative addressing method is exploited [22]. The contributions of this paper are given below:

- Devising a row-balanced dual-ratio sparsity algorithm for improving the accuracy of the LSTM models while considering the hardware implementation (BRDS algorithm).
- Presenting a low-energy yet high-speed FPGA-based hardware accelerator based on the above pruning algorithm for facilitating the implementations of BRDS-based sparse models.

The remainder of the paper is organized as follows. Section 2 provides basic concepts of LSTM as well as a review of prior work on FPGA-based LSTM architectures. The proposed row-balanced dual-ratio sparsity algorithm is presented in Section 3. Section 4 provides the details of the proposed hardware accelerator. In Section 5, the efficacies of the proposed algorithm and accelerator are evaluated, and finally, the paper is concluded in Section 6.

## 2 LSTM Basic Concepts and Related Work

In this section, first, the internal structure of the considered LSTM network layer is described and then prior work dealing with the LSTM implementation on FPGA platforms and several LSTM pruning and compression algorithms are briefly reviewed.

### 2.1 Considered LSTM Network

In this work, we use the LSTM network of [7], which has a simple structure with an acceptable output accuracy. A layer of this network consists of the cells (to store prior information) and gates (*i.e.*, $f_t$, $i_t$, $g_t$ and $o_t$) to control whether to remember or forget prior information (*i.e.*, $c_{t-1}$), inputs (*i.e.*, $x_t$), and outputs of the previous time step (*i.e.*, $h_{t-1}$). Based on this, an LSTM layer may be described as [7]:

$$\begin{aligned} f_t &= sig(W_{fx}x_t + W_{fh}h_{t-1} + b_f) \\ i_t &= sig(W_{ix}x_t + W_{ih}h_{t-1} + b_i) \\ g_t &= \tanh(W_{gx}x_t + W_{gh}h_{t-1} + b_g) \\ o_t &= sig(W_{ox}x_t + W_{oh}h_{t-1} + b_o) \end{aligned} \quad (1)$$

$$\begin{aligned} c_t &= f_t \odot c_{t-1} + i_t \odot g_t \\ h_t &= o_t \odot tanh(c_t) \end{aligned} \quad (2)$$

where $tanh$ and $sig$ (*i.e.*, Sigmoid) are logistic activation functions and $\odot$ is the dot product operation. Also, $W$ and $b$ denote the weight matrix and bias vector, respectively. Gates $f$, $i$, $g$, and $o$ correspond to the forget, input, candidate cell, and output gates, respectively. Weight matrices (*i.e.*, $W_x$ and $W_h$) and the bias vector are determined for each gate (*e.g.*, weight matrices and bias vector for gate $f$ are denoted as $W_{fx}$, $W_{fh}$, and $b_f$, respectively). In addition, $t$ denotes the current time step. The size of the vector $x$ is $X$ when sizes of the vectors $h$, $b$, and $c$ are $H$. Similarly, sizes of matrices $W_x$ and $W_h$ are $H \times X$ and $H \times H$, respectively. The activation functions perform element-wise computations on their input vectors. The internal structure of the considered LSTM layer is shown in Fig. 1.

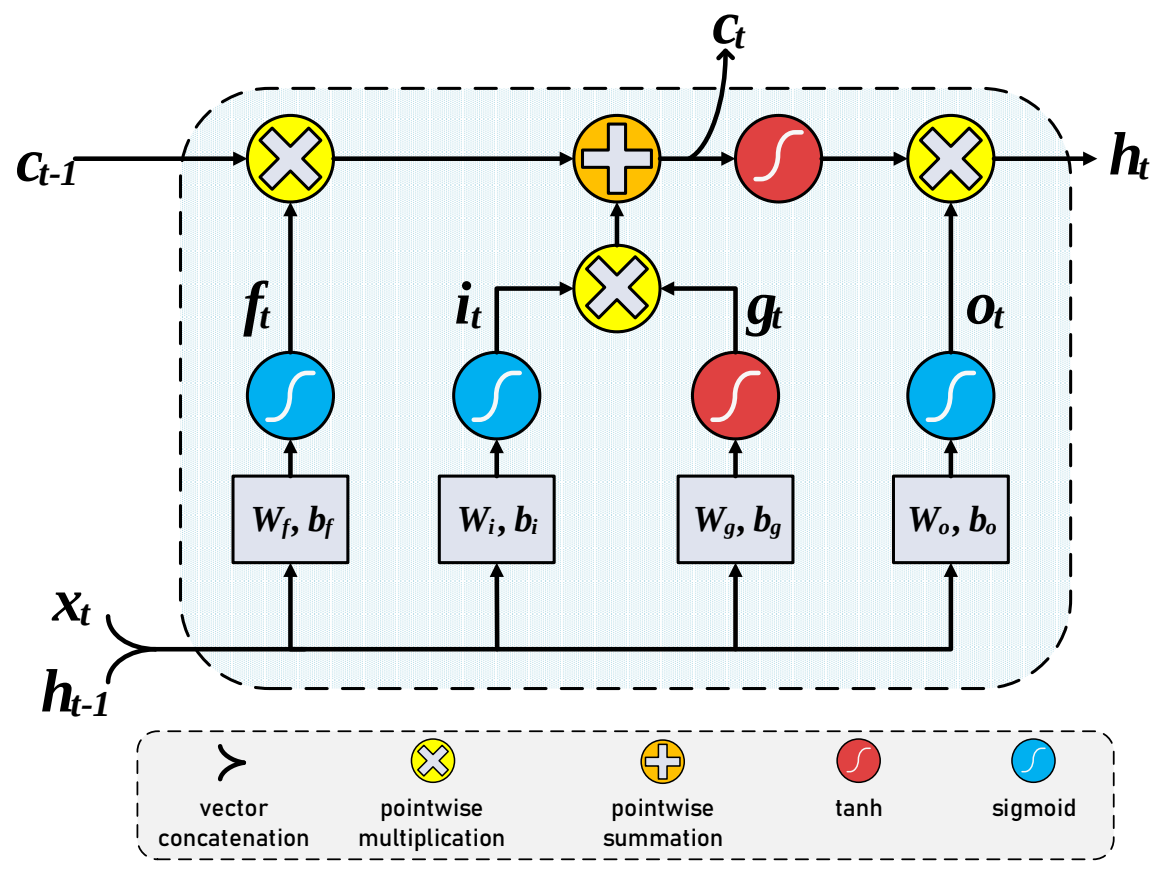

Fig. 1. The internal structure of the considered LSTM layer.

### 2.2 Related Work

The compression of the network model could lead to speed and energy efficiency improvements of the inference phase [10]. The improvements are achieved by reducing the memory usage and bandwidth and the computational requirements of an NN-based inference. Well-known compression techniques consist of pruning [11], sparsity-inducing regularization [12], and quantization [4]. Several structured sparsity methods have been proposed in prior studies. The proposed algorithms considered constraints on the locality of non-zero weights to limit the scattering of the zero weights in the weight metrices [8], [13]. Compared to unstructured sparsity, accelerating the structured sparsity using special hardware is more feasible and affordable.

The proposed LSTM architecture in [4] utilized weight compression and pruning techniques to increase speed and energy efficiency. The gains (in terms of energy efficiency and computational speed) were obtained at the cost of considerable hardware resource usage. The method proposed in [14] reduced the LSTM network size and controlled the network irregularity. It made use of block-circulant matrices [15] (*i.e.*, arbitrary size circulant submatrices), and further applying the FFT algorithm to accelerate the compute-intensive circulant convolution operations. Variable submatrix sizes provided a tradeoff between the compression ratio and the accuracy degradation.

In [16], first, an algorithm for reducing the computations of the Gated Recurrent Unit (GRU) network was suggested. The algorithm induced sparsity in the inputs and activations, thereby lowering the computations. Next, an accelerator architecture, called DeltaRNN, which skipped the updating of an RNN when the input changes were below a certain threshold, was presented. In [9], *Bank-Balanced Sparsity* (BBS), which partitions each weight matrix row into banks for parallel computing, was proposed. This method adopted fine-grained pruning inside each bank to maintain the model accuracy. The architecture, which was fully parallel, had the drawback of considering the same

PEs for the forward and recurrent weights.

The proposed BRDS hardware architecture, which is an extension of the POLAR accelerator in [7] designed for the inference phase of dense networks, has the ability to support sparse LSTM networks. By utilizing parallel modules and an addressing technique, performing sparse operations efficiently as well as higher efficacy for BRDS compared to POLAR were achieved.

# 3 Row-Balanced Dual-Ratio Sparsity Pruning

## 3.1 Row-Balanced Pruning of Weight Matrices

Fig. 2 illustrates the original and pruned matrices with the sparsity ratio of 50%. Fine-grained pruning simply omits the smallest 50% of the weights globally which leads to unstructured sparse matrix (Fig. 2(b)). Block sparsity induces a block sparse matrix (Fig. 2(c)) by setting the block size to $m{\times}m$ (which in this example is 2×2) and the block representative (as the metric for pruning the blocks) with the block average. Bank-balanced pruning induces a bank-balanced sparse matrix (Fig. 2(d)) by splitting each row into two equal-sized banks and applying fine-grained pruning inside each bank independently.

In this work, we propose row-balanced sparsity whereby the same number of elements from every row of a given weight matrix are pruned. The row-balanced sparse matrix of Fig. 2(a) is shown in Fig. 2(e) where the smallest 50% elements in each row have been removed. The pseudo-code of the row-balanced sparsity is shown in Fig. 3. The inputs to this algorithm are the weight matrix and the expected sparsity ratio while the output is the pruned matrix. It prunes each row separately based on the given sparsity ratio. To do this, based on the defined sparsity ratio (*i.e.*, $Spar$%), some of the elements are pruned (in order from the smallest to the largest values).

## 3.2 Row-Balanced Dual-Ratio Sparification Algorithm

As mentioned in subsection 3.1, in LSTM networks, for two different sets of weights (*i.e.*, $W_x$ and $W_h$), the proposed accelerator can consider two different sparsity ratios. The sparsity ratios are denoted by $Spar_h$ and $Spar_x$ for $W_h$ and $W_x$, respectively. Applying the aforesaid row-balanced sparsity approach, every feed-forward (recurrent) weight matrix, *i.e.*, $W_{fx}$, $W_{ix}$, $W_{gx}$, and $W_{ox}$ ($W_{fh}$, $W_{ih}$, $W_{gh}$, and $W_{oh}$) will have the same numbers of non-zero elements per row. Choosing different sparsity ratios for feed-forward and recurrent weights alleviates the accuracy degradation resulting from the pruning process. This originates from the fact that the algorithm decides which weights are more important to keep. As an example, in Fig. 4, the effect of having two different sparsity ratios on the accuracy of an LSTM model is shown. The dataset here is PTB [17] with an input size of 1,500. For this example, the overall sparsity ($OS$) of 65% is considered. When $Spar_x$ and $Spar_h$ were both set to 65%, the perplexity (the metric widely used in NLP) became large while the best perplexity was achieved when we set $Spar_h = 60\%$ and $Spar_x = 70\%$. A low value for the perplexity shows a well-trained LSTM network [20].

| | | | | | | | |
|---|---|---|---|---|---|---|---|
| 0.3 | 0.1 | 0.4 | -0.5 | 0.1 | -0.1 | 0.2 | 0.6 |
| 0.3 | 0.4 | 0.6 | 0.1 | -0.1 | 0.2 | 0.5 | -0.5 |
| 0.1 | 0.4 | -0.2 | 0.5 | -0.2 | 0.5 | 0.3 | -0.4 |
| 0.2 | -0.6 | 0.6 | 0.5 | 0.1 | 0.2 | 0.4 | 0.7 |

(a) Original dense matrix

| | | | | | | | |
|---|---|---|---|---|---|---|---|
| | | 0.4 | -0.5 | | | 0.2 | 0.6 |
| | 0.4 | 0.6 | | | | 0.5 | -0.5 |
| | 0.4 | | 0.5 | | 0.5 | | -0.4 |
| | -0.6 | 0.6 | 0.5 | | | 0.4 | 0.7 |

(b) Unstructured sparse matrix by global pruning

| | | | | | | | |
|---|---|---|---|---|---|---|---|
| 0.3 | 0.1 | | | | | 0.2 | 0.6 |
| 0.3 | 0.4 | | | | | 0.5 | -0.5 |
| | | -0.2 | 0.5 | | | 0.3 | -0.4 |
| | | 0.6 | 0.5 | | | 0.4 | 0.7 |

(c) Block sparse matrix by pruning 2x2 blocks according to block average

| | | | | | | | |
|---|---|---|---|---|---|---|---|
| | | 0.4 | -0.5 | | | 0.2 | 0.6 |
| | 0.4 | 0.6 | | | | 0.5 | -0.5 |
| | 0.4 | | 0.5 | | 0.5 | | -0.4 |
| | -0.6 | 0.6 | | | | 0.4 | 0.7 |

(d) Bank-balanced sparse matrix by local pruning inside each 1x4 bank

| | | | | | | | |
|---|---|---|---|---|---|---|---|
| 0.3 | | 0.4 | -0.5 | | | | 0.6 |
| | 0.4 | 0.6 | | | | 0.5 | -0.5 |
| | 0.4 | | 0.5 | | 0.5 | | -0.4 |
| | -0.6 | 0.6 | 0.5 | | | | 0.7 |

(e) Row-balanced sparse matrix

Fig. 2. Comparing different pruning methods with Row-Balanced.

One may find more information about the perplexity metric in [21].

Similar to the previous pruning methods (see, *e.g.*, [4], [9]), we apply the row-balanced pruning method iteratively to a pre-trained network and retrain the network after each pruning iteration to partially restore the model accuracy. Since there are two matrices which should be pruned simultaneously with different sparsity ratios and the sensitivities of the output quality to the sparsity ratio are different for the two matrices, we determine the sparsity ratios considering an overall sparsity target provided by the designer. The (minimum) sparsity target is determined based on the number of weights that the designer wants to store in the on-chip memories. The goal, therefore, is to achieve the best model accuracy given a designer-specified lower bound on the sparsity factor. Since the value of $Spar_x$ and $Spar_h$ cannot be directly obtained from the value of $OS$ due to the their dependency on the dataset, these values should be determined by exploring different possible combinations of them as shown in Fig. 4.

Based on the above discussion, a heuristic algorithm for pruning the $W_x$ and $W_h$ weight matrices is presented. The pseudo-code of the pruning method which is inducing row-balanced dual-ratio sparsity is shown in Fig. 5. It iteratively explores the search space to find the best sparsity ratios (*i.e.*, the best values for $Spar_h$ and $Spar_x$). In each pruning iteration, based on the considered sparsity ratios, the weights with small importance are dropped. In this algorithm, the importance of weights is represented by their internal ranking within the row, which is dictated by their absolute values. To reduce the accuracy loss due to pruning, in each iteration, the pruned network is retrained to

**Input:**
The matrix to be pruned, $W$;
The expected sparsity, $Spar$%;
**Output:**
The pruned matrix, $W_p$;
**for** each $W_i \in W.rows$ **do**
    Prune the smallest $Spar$% of $W_i$;
**end for**
**return** the pruned matrix, $W_p$;

Fig. 3. The Row-Balanced Pruning Algorithm.

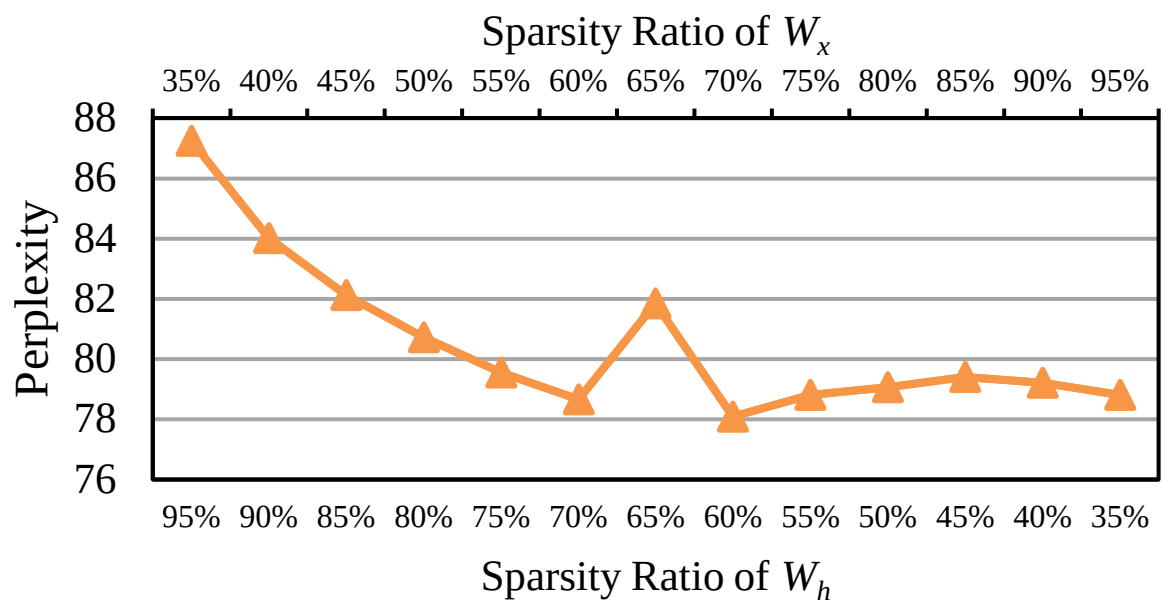

Fig. 4. The effect of Dual-Ratio Sparsity on the perplexity of PTB dataset.

determine the corresponding accuracy of the chosen sparsity ratios. For retraining, we freeze the weights that are set to zero (*i.e.*, the dropped ones) and tune the other network weights.

In the proposed algorithm, to lower the accuracy loss due to the pruning, in the first step, we suggest increasing the pruning ratios (*i.e.*, $Spar_h$ and $Spar_x$) gradually with the same step size ($\alpha$) from zero to the predefined overall sparsity ($OS$). The pruned network at this point is considered as the initial point for searching. We denote it as $NN_{P,I}$. Next, to explore the search space, one of the sparsity ratios (*e.g.*, $Spar_h$) is increased by a predefined step ($\delta_h$) while the other one (*e.g.*, $Spar_x$) is decreased by its predefined step ($\delta_x$). In each iteration, the chosen tuple of the sparsity is applied on the pruned network of the previous iteration. Altering the sparsity ratios is continued till one of them reaches 0 or 100%. Next, this process is repeated again (by

**Input:**
The weights of the LSTM layer to be pruned, $W_x$ and $W_h$;
The expected overall sparsity, $OS$;
**Output:**
The maximum model accuracy, $MA$;
The sparsity ratios with the maximum model accuracy $Spar_{x,MA}$ and $Spar_{h,MA}$;
1: set $Spar_x$ and $Spar_h$ to 0;
2: **while** $Spar_x < OS$ and $Spar_h < OS$ **do**
3: Increase $Spar_x$ and $Spar_h$ by $\alpha$;
4: Prune $W_x$ and $W_h$;
5: Retrain the network;
6: Save the pruned network as $NN_{P,I}$;
7: **while** $Spar_x < 100\%$ and $Spar_h > 0\%$ **do**
8: Increase $Spar_x$ by $\delta_x$;
9: Decrease $Spar_h$ by $\delta_h$;
10: Prune $W_x$ and $W_h$;
11: Retrain the network and save model accuracy to $Acc$;
12: **if** $Acc > MA$ **do**
13: $MA = Acc$;
14: $(Spar_{x,MA}, Spar_{h,MA}) = (Spar_x, Spar_h)$;
15: Load the pruned network $NN_{P,I}$;
16: **while** $Spar_x > 0\%$ and $Spar_h < 100\%$ **do**
17: Decrease $Spar_x$ by $\delta_x$;
18: Increase $Spar_h$ by $\delta_h$;
19: Prune $W_x$ and $W_h$;
20: Retrain the network and save model accuracy to $Acc$;
21: **if** $Acc > MA$ **do**
22: $MA = Acc$;
23: $(Spar_{x,MA}, Spar_{h,MA}) = (Spar_x, Spar_h)$;
24: **return** $MA$, $(Spar_{x,MA}, Spar_{h,MA})$;

Fig. 5. The BRDS Algorithm.

starting from the $NN_{P,I}$) considering the opposite direction for the sparsity ratios. For each chosen tuple of the sparsity ratios, the accuracy of the network is determined. At the end, the algorithm returns the best tuple. To generate the sparsity ratios with the maximum model accuracy ($Spar_{x,MA}$ and $Spar_{h,MA}$), the BRDS algorithm is executed only once. The pruned network is used multiple times, so the inference takes a long time which amortizes the cost of the retraining algorithm.

In addition, the execution time of the algorithm depends on $OS$, $\alpha$, $\delta_x$, $\delta_h$, $ept$ (the time needed for each epoch), and $n_{re}$ (the number of epochs needed for the retraining. The formulas below can be used to attain the execution time of the algorithm assuming the pretrained network is available. The parameter $ept$ depends on both the size of the model that the user is trying to prune and the hardware that is going to be utilized to perform the algorithm. The parameters $ex_1$, $ex_2$, $ex_3$, and $ex_{tot}$ show the execution time of lines 1-6, 7-14, 15-24, and the whole algorithm, respectively.

$$ex_1 = \frac{OS}{\alpha} \times ept \times n_{re} \quad (3)$$

$$ex_2 = \min\left(\frac{100 - OS}{\delta_x}, \frac{OS}{\delta_h}\right) \times ept \times n_{re} \quad (4)$$

$$ex_3 = \min\left(\frac{100 - OS}{\delta_h}, \frac{OS}{\delta_x}\right) \times ept \times n_{re} \quad (5)$$

$$ex_{tot} = ex_1 + ex_2 + ex_3 \quad (6)$$

## 4 Hardware Architecture

As discussed before, unstructured sparsity leads to unbalanced computations as well as irregular memory accesses. To take advantage of the structured sparsity introduced by the BRDS algorithm, an efficient LSTM hardware accelerator, called BRDS LSTM accelerator, is presented next. The internal structure of the accelerator, which is shown in Fig. 6, is based on POLAR accelerator of [7] with some modifications to support dual sparsity.

The BRDS accelerator consists of seven main modules including *DRAM Controller*, *Embedded Memory*, *Address Decoder*, *Gate*, *Function*, *Buffer*, and *LSTM Controller*. The bit width of the datapath is $n$, and the data is represented in fixed-point two's complement $n$-bit binary format. *DRAM Controller* performs load and store instructions related to off-chip DRAM. A load instruction may occur when data should be read from off-chip DRAM and written onto On-chip memories. Similarly, when outputs are ready, *DRAM Controller* should read the data and store them on the off-chip DRAM. It should be mentioned that most of the times, the weights can be fully fit in FPGA embedded memories. In the cases where, even with pruning, we cannot fit the data in the FPGA memories, the proposed accelerator utilizes this module as the interface to off-chip DRAM. Note that only non-zero elements of the sparse matrices are fetched consecutively to efficiently utilize the off-chip memory bandwidth.

The module *Gate* (see Fig. 6) includes two Mult Arrays (MAs) working concurrently, an MA Selector, a Tree Adder, and an Accumulator. Since there are two different pruning ratios ($Spar_h$ and $Spar_x$), we consider two MAs with different sizes, named Small and Large. By using a multiplexer in the MA Selector, the accelerator can choose, for each set of weights, which MA to be employed. The number of elements in each row of $W_h$ and $W_x$ which are $H$ and $X$, respectively, become $H_{SP}$ and $X_{SP}$, after running the BRDS algorithm. The weight matrix with a larger (smaller) number of elements in each row (*i.e.*, $H_{SP}$ or $X_{SP}$) utilizes Large (Small) MA. Two MAs working together conduct $R$ signed multiplications in parallel. Large (Small) MA includes a Mult Array component which conducts $R_L$ ($R_S$) parallel $n$-bit multiply operations in a single cycle ($R = R_L + R_S$). It should be noted that the parameters $R_L$ and $R_S$ show the level of parallelization for each weight matrix (*i.e.*, $W_h$ and $W_x$). Here, Large (Small) MA processes the weights with the larger (smaller) number of elements in each row. To fully utilize the MAs, the BRDS accelerator considers $R_L$ and $R_S$ in a way that $R_S/R_L$ be equal to $(\min\{X_{SP},\ H_{SP}\})/(\max\{X_{SP},\ H_{SP}\})$. In this way, the ratio of the number of non-zero elements in small and large matrices would be equal to the ratio of the number of multipliers in small and large MAs. Therefore, the number of clock cycles needed for processing each weight matrix is the same, and there is no time in which one MA is not utilized.

It is worth mentioning that the parameter $R$, which shows the number of parallel multiplication operations for every row, determines the latency and the resource usage. Since, in the BRDS hardware, the input rows are pruned, to reach a higher level of parallelism, we propose a new parallelization factor, called $Q$, showing the number of rows whose corresponding calculations could be performed parallel. Therefore, parameter $Q$ shows the number of modules *Gate* (*Buffer* and *Function* as well) working in parallel. The outputs of the Large and Small MAs are concatenated and passed onto the Add Array component in the module *Gate*. The Add Array component utilizes a tree of $n$-bit adders, which gives the summation of its $R$ input operands. To perform the additions and multiplications, we use DSP blocks in the FPGA. Due to the architecture of the DSP blocks, we are able to perform some functions together. In the Tree Adder component, we use three-input adders where it is possible to minimize resource utilization. The internal structure of a DSP block of the Xilinx FPGAs (DSPE48) is shown in Fig. 7. The specified path is utilized for realizing the three-input adders. The current output of this component is added to its previous one in an Accumulate component that is implemented by a DSP block. The output of the Accumulate component is passed to an Add unit to take biases into account. All these components as one unit perform the computation of the MxV.

The proposed accelerator truncates the output of each add and multiply unit to $n$ bits. To alleviate the impact of overflow in the result, we utilize Recovery units after each Add and Multiply unit suggested in [7]. The module *Function* (see Fig. 6) performs operations that are pointwise (*i.e.*, *sig*, *tanh* and (2)). This module generates the output $h$ and the cell state $c$, which are written to their corresponding space in the module *Embedded Memory*. The operations of this module are overlapped by those of the module *Gate* where this overlap is provided by the module *Buffer*. The proposed accelerator utilizes piecewise linear approximation of the activation functions (*e.g.*, *sigmoid* ($\sigma$) and *tanh*) to balance speed, accuracy, area, and power consumption. In each piece, two $n$-bit coefficients $a$ and $b$ are obtained and stored in LUTs. Hence, the word size of the LUTs are $2n$ bits. In the activation function component, by employing the add and multiply operations in one DSP block, the output ($a \times x + b$) of the activation functions are determined. The operations of (2) are done by deploying a multiply unit and an add unit in the module *Function*. Because the output of the multiply unit should be passed to the module *Embedded Memory*, these units are implemented separately by DSP blocks.

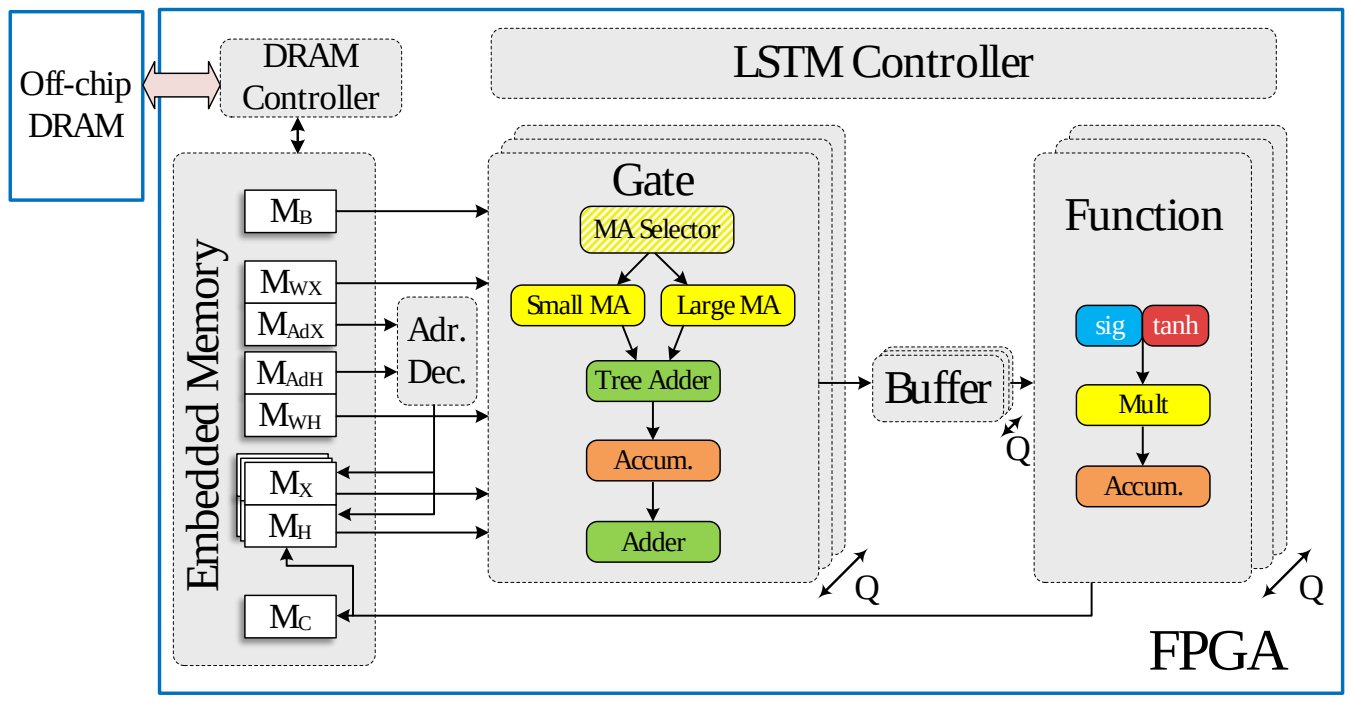

Fig. 6. The internal structure of the proposed BRDS LSTM accelerator.

To transfer the data from the module *Gate* to the module *Function*, and also to feed back the result of the module *Function* to itself, the module *Buffer* (see Fig. 6) is deployed and used with the same approach as that of the POLAR architecture [7]. The module *Embedded Memory* (see Fig. 6) stores the weights, biases, inputs, relative addresses of the inputs, cells, and outputs in the embedded memory banks of the FPGA. To store the weights, two memory arrays denoted by MWX and MWH are employed. Only the nonzero elements of the matrices $W_{fx}$, $W_{ix}$, $W_{gx}$, and $W_{ox}$ are stored in MWX. We use relative row index and cumulative pointer to store sparse matrices. The relative row index for each element shows the number of zero elements before it. Each $R_X$ ($R_H$) nonzero elements of $W_x$ ($W_h$) weights are stored in four consecutive rows. Hence, the size of MWX is $4H \times X_{SP} \times n$ bits where the width of each row of this memory is $R_X \times n$. Similarly, the elements of matrices $W_{fh}$, $W_{ih}$, $W_{gh}$, and $W_{oh}$ are stored in MWH. The size of MWH is $4H \times H_{SP} \times n$ bits, where the width of each row of this memory is $R_H \times n$.

The memory array $M_B$ of the module *Embedded Memory* is used to store the biases. The size of $M_B$ is $4H \times n$ bits with a word-width of $n$ bits. The $i^{\text{th}}$ row of the biases $b_f$, $b_i$, $b_g$, and $b_o$ are stored in the four consecutive rows of the memory $M_B$. The memory array $M_X$ stores $N$ time steps of the input dataset where the size of the input dataset, in each time step, is $X \times n$ bits. To gain more throughput, for the inputs, we use duplicate memories in parallel. Because of the utilization of the Dual Port RAMs, we need $R_X/2$ BRAMs ($M_X$) for storing the inputs. In this work, for simplicity, we consider a single time step ($N = 1$). The memory array $M_H$ stores the outputs of the current and previous time steps ($h_t$ and $h_{t-1}$) with the size of $H \times n$ bits. Similar to the input memory array, duplicated memories for the outputs were utilized ($R_H/2$ BRAMs for $M_H$). At the beginning of each time step, this memory contains $h_{t-1}$. After generating each element of $h_t$, this element will be stored on their corresponding rows in all replicated memories of $M_H$. By using duplicate memories, the proposed architecture will have a much better performance by increasing the throughput at the cost of using a reasonable amount of embedded memory. The generated cells and the ones for the previous time step are accessible from the memory array $M_C$. The word size of this memory is $n$ bits while its size is $H \times n$ bits. In this memory approach, before replacing previous $c_t$ (*i.e.*, current $c_{t-1}$), the cell is fetched and then replaced with the current $c_t$.

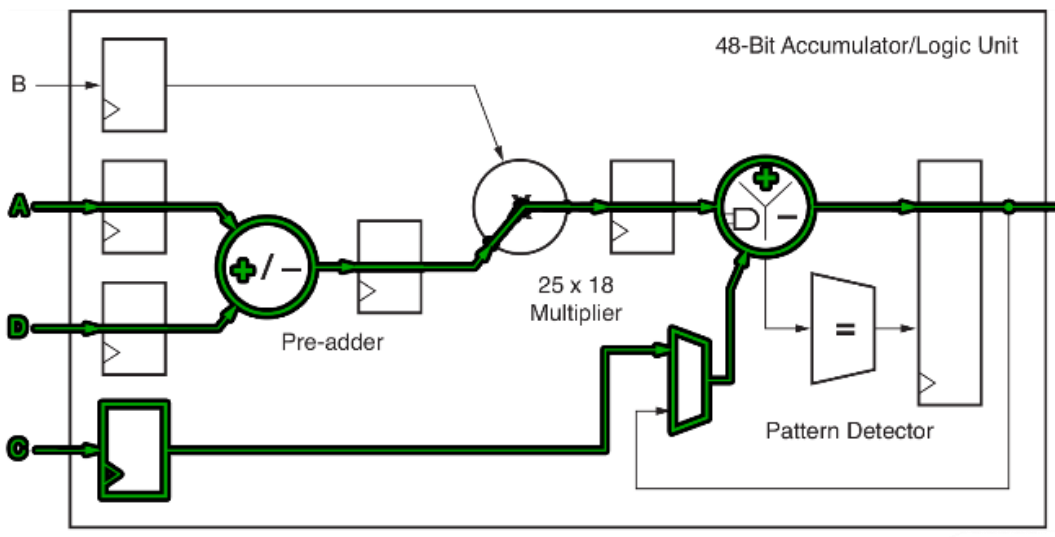

Fig. 7. The internal structure of a DSP block configured for three-input adder.

We store relative addresses corresponding to the memories MWX and MWH in memories MAdX and MAdH, respectively. The module *Address Decoder*, which consists of small add units, decodes the relative addresses to obtain cumulative pointer addresses. The pattern of storing $W_h$ in MWH and their corresponding relative addresses in MAdH are illustrated in Fig. 8. In this example, $H$, $Spar_h$, and $R_H$ (the parallelization factor of $W_h$) are considered 4, 50%, and 2, respectively. Storing $W_x$ and its relative addresses in the corresponding memories are performed similar to that of $W_h$.

Based on the proposed datapath for the accelerator architecture, a designer may control the trade-off between the resource usage and the latency of the architecture simply by adjusting the parameters $R$ and $Q$ at the design time. To switch from one parallelization factor to another, one needs to change the number of mult and add units in MAs and Add Array components, respectively. Also, the designer should change the size of the delay units in the module *Buffer*. It is worth mentioning that if the parallelization factor $R_x$ ($R_h$) is chosen greater than $X_{SP}$ ($H_{SP}$), the designer should use the parameter $Q$ and utilize multiple number of modules *Gate*, *Buffer*, and *Function*.

The module *LSTM Controller* in the proposed architecture performs the control of the complicated timing scheme of the LSTM network. This module generates proper signals with proper timing to meet the architecture requirements. Details of the timing of this architecture is the same as that of the POLAR architecture [7].

## 5 Results and Discussion

In this section, the accuracy of the proposed pruning algorithm is evaluated by applying it to some LSTM networks. Also, the design parameters of the proposed accelerator, as well as its efficacy compared to several prior works, are assessed by implementing the accelerator on an FPGA.

### 5.1 Model Accuracy of BRDS Algorithm

To evaluate the accuracy of the BRDS pruning algorithm, the algorithm was applied to an LSTM language model of the PTB dataset [17], the IMDB Movie Reviews dataset [18], and the TIMIT dataset [19]. The PTB dataset, widely used in NLP researches, includes 929K training, 73K validation, and 82K test words. The IMDB dataset has 50,000 highly polar positive and negative movie reviews for binary sentiment classification. It includes 25,000 reviews for training and 25,000 reviews for testing. The TIMIT dataset has been provided for the study of acoustic-phonetics. It includes recordings of 630 speakers of eight major dialects of American English. For the LSTM speech recognition model, we set the input size to 153 and the hidden state to 1024 which are the same as the prior studies ([4], [9]).

In this study, the accuracy of the BRDS is compared to three prior pruning approaches including unstructured sparsity, block sparsity, and bank-balanced sparsity (BBS) [9]. For the studies of this section, 64 banks in the case of BBS method and 4×4 blocks in the case of block sparsity method were utilized which are the same sizes as those used for the hardware implementation in [9]. The trade-offs between the sparsity ratio and the accuracy of different sparsity patterns on PTB, TIMIT, and IMDB datasets are depicted in Fig. 9. To perform experiments on the LSTM language model, we employed the large model of PTB dataset with 1,500 inputs. Also, the perplexity metric, as a widely used merit parameter in quantifying language model quality [9], was exploited as the error metric for this dataset. As the results in Fig. 9(a) show, the perplexities of the suggested pruned network by the BRDS algorithm is lower than the other pruning approaches. Also, the BRDS algorithm preserved the perplexity even until 85% of the weights were pruned. On average, the proposed algorithm led to 0.7% lower perplexity compared to that of BBS method for the pruning ratio ranging from 0 to 90% with the interval of 5%.

The LSTM model chosen for the experiments on TIMIT dataset was also the same as the prior works ([4], [9]). The metric for evaluating the accuracy of the acoustic model is PER (Phone Error Rate) which is a merit evaluation parameter for speech recognition models [9]. The results in Fig. 9(b) show that the PER of the network pruned by the BRDS algorithm is, on average about 0.1%, less than that of the BBS method. As the results in Fig. 9(c) show, our method outperforms other algorithms in almost all the cases, particularly in larger sparsity ratios. Compared to the BBS sparsity method, the proposed pruning algorithm resulted in 0.7% lower accuracy loss. Note that the efficacy of the proposed pruning algorithm is reduced in the case of this dataset compared to the other considered datasets due to its small size.

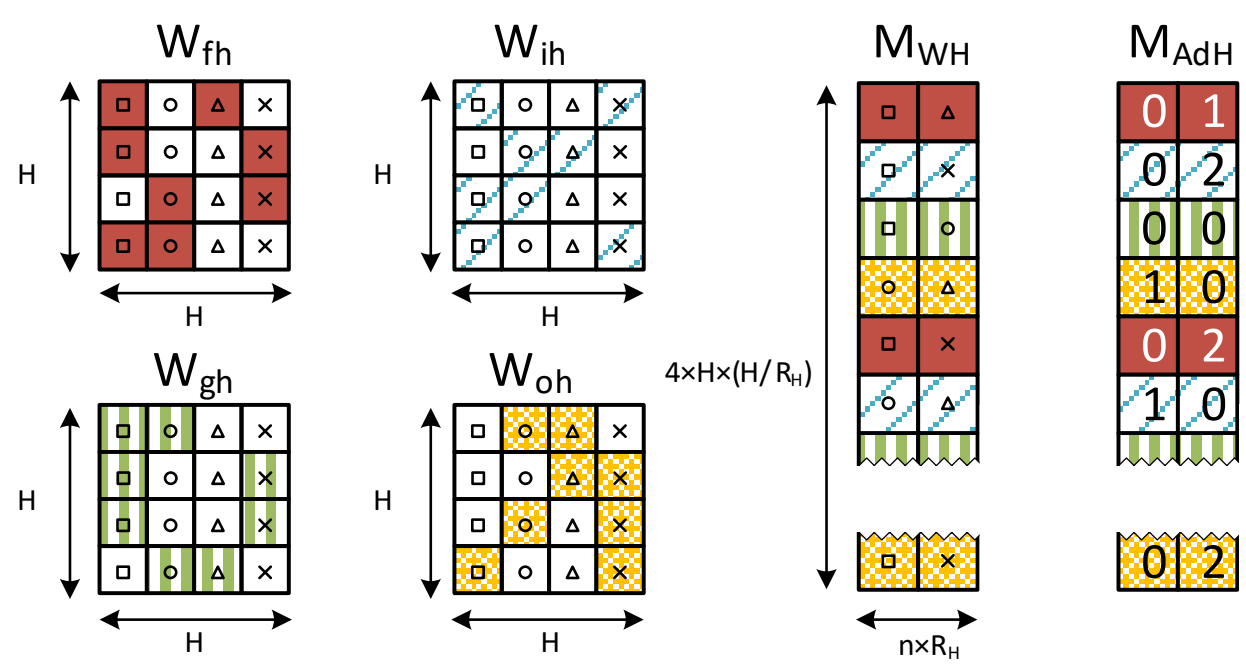

Fig. 8. Pattern of storing $W_h$ and its relative addresses in $M_{WH}$ and $M_{AdH}$. The elements of each row are distinguished by the square, circle, triangle, and cross shapes.

### 5.2 Efficiency of BRDS Accelerator

To evaluate the efficiency of the architecture of the proposed accelerator, it was implemented on FPGA for executing the TIMIT dataset with the same configurations provided in [4], [9], [14]. The design parameters of the BRDS are compared with four state-of-the-art works including the ones proposed in [4], [9], [14], [16]. The focus of all of these prior works were on implementing the LSTM networks on FPGAs using weight pruning and compression. The work in [16] was evaluated on GRU which is simpler than LSTMs in terms of its computational complexity. Thus, the reported design parameters of [16] should be looked at as optimistic values when compared to those of the LSTM accelerators. This work used the delta network algorithm to reduce MxV operations and skipping unimportant cell activation changes (the changes were below a threshold value) to reduce memory accesses. To perform a better comparison, we implemented the BRDS design on an FPGA (XCKU9P) with the same family as that of [4] by exploiting Xilinx VIVADO 2018.2 tool.

In this study, the pruning ratio was set to 87.5% (the same as [9], [14]). Also, since the data bit width was considered as 16 bits in most of the prior architectures (*e.g.*, [9], [14], [16]), without loss of generality, the same size was considered for the BRDS accelerator. For pruning the network, we applied the BRDS algorithm for the overall sparsity ($OS$) of 87.5% (the same sparsity ratio as most of the prior works) on the TIMIT dataset. The best $Spar_h$ and $Spar_x$ given by the algorithm were 87.5% for both. It is worth mentioning that because parameters $X$ and $H$ were different in this design, having the same sparsity ratios for $W_h$ and $W_x$ did not mean the same number of elements in each row for them. Thus, even by having the same sparsity ratios, the numbers of elements to be pruned were different for $W_h$ and $W_x$. After running the BRDS pruning algorithm, the parameters $X_{SP}$ and $H_{SP}$ were 20 and 64, respectively.

As mentioned in Section 4, to fully utilize the MAs in the BRDS accelerator, the parameters $R_L$ and $R_S$ were considered such that $R_S/R_L$ be equal to $(min\{X_{SP}, H_{SP}\})/(\max\{X_{SP}, H_{SP}\})$. Therefore, we considered $R_S/R_L$ as 80/256 making the parallelization factors $R$ and $Q$ equal to 336 and 4, respectively. TABLE 1 shows the resource utilization of the BRDS accelerator with the mentioned configuration on the XCKU9P Xilinx FPGA device. Also, TABLE 2 shows the frequency, sparsity ratio, accuracy degradation, GOPS, power, GOPS/W, and effective GOPS, GOPS/W, and DSP and logic efficiency of the BRDS accelerator and those of the considered prior works. The Effective Throughput (GOPS) is defined as $(GOPS)/(1-sparsity)$ which takes the impact of pruning into account

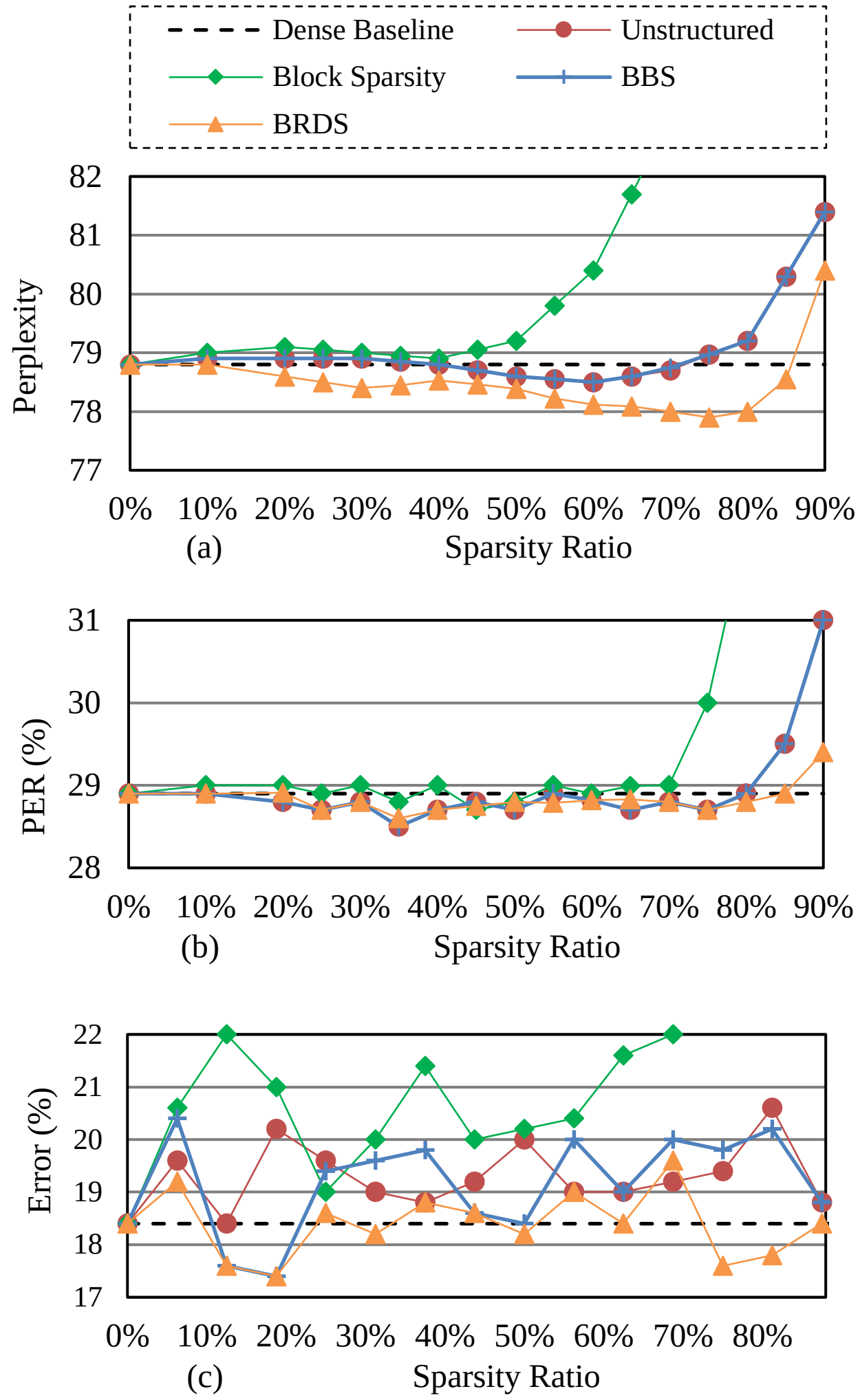

Fig. 9. The accuracy-sparsity tradeoff on (a) PTB, (b) TIMIT, and (c) IMDB datasets.

[9]. In this work, although the operating frequency of the BRDS architecture could be increased to 238MHz, it was set to 200MHz to have a fair comparison with the selected prior works. The design parameters of the prior works were borrowed from their published papers. As the results in TABLE 2 show, the throughput (GOPS) of the BRDS accelerator is higher (up to 52.7%) than that of [14], [16] while smaller (up to 52%) than those of [4], [9]. The power consumption was extracted using Xilinx Power Estimator. The switching activity was set by the tool based on inputs and weights for the TIMIT dataset stored in the embedded memories. The power consumption of the BRDS accelerator is much less than that of other selected works except for [16], mostly due to its smaller operating frequency. Additionally, the GOPS/W of the BRDS shows a higher number compared to those of other works except for [16] which has

TABLE 1
RESOURCE UTILIZATION OF THE BRDS ACCELERATOR IMPLEMENTED ON XCKU9P FOR TIMIT DATASET ($OS$ = 87.5%).

| | LUT | FF | DSP | BRAM |
|---|---|---|---|---|
| Available | 274080 | 548160 | 2520 | 912 |
| Utilization | 5600 | 83710 | 1600 | 724 |
| Utilization (%) | 2 | 15.3 | 63.5 | 79.4 |

TABLE 2
COMPARISON OF THE DESIGN PARAMETERS OF DIFFERENT STATE-OF-THE-ART LSTM ACCELERATORS.

| | ESE [4] | C-LSTM [14] | DeltaRNN [16] | BBS [9] | BRDS |
|---|---|---|---|---|---|
| Platform | XCKU060 | Virtex-7 | XC7Z100 | Arria 10 GX1150 | XCKU9P |
| Frequency (MHz) | 200 | 200 | 125 | 200 | 200 |
| Sparsity (%) | 88.7 | 87.5 | - | 87.5 | 87.5 |
| Quantization | fixed-12 | fixed-16 | fixed-16 | fixed-16 | fixed-16 |
| Accuracy Degradation | 0.30% | 0.32% | - | 0.25% | 0.25% |
| Throughput (GOPS) | 282 | 131 | 192 | **304** | 200 |
| Power (W) | 41.0 | 22.0 | **7.3** | 19.1 | 9.0 |
| Energy Efficiency (GOPS/W) | 6.9 | 6.0 | **26.3** | 15.9 | 22.2 |
| Effective Throughput (GOPS) | **2497** | 1049 | 1198 | 2433 | 1600 |
| Effective Energy Efficiency (GOPS/W) | 60.9 | 47.7 | 163.3 | 127.4 | **177.8** |
| Effective DSP Efficiency (GOPS/#DSP) | 1.66 | **0.39** | 1.56 | 1.60 | 1.00 |
| Effective Logic Efficiency (GOPS/#Cell) | 0.008 | 0.002 | 0.005 | 0.008 | **0.286** |

a slightly better energy efficiency. The effective GOPS (GOPS with considering the sparsity ratio) of the BRDS is, on average, about 43% higher than those of [14], [16] while it is on average about 54% lower than those of the [4], [9]. Moreover, the BRDS accelerator outperforms all of the other works in terms of effective GOPS/W. The effective GOPS/W of the BRDS, on average (up to), is 2.3× (3.7×) higher than those of the other selected works. Finally, GOPS/#DSP and GOPS/#Cell are considered to normalize the effective throughput based on the amount of utilized DSP and logic cell (*i.e.*, ALM for Intel and LUT for Xilinx devices).

## 6. CONCLUSION

In this paper, first, the BRDS, a row-balanced dual-ratio sparsity algorithm, was presented to improve the accuracy of LSTM models considering their hardware implementation. Additionally, BRDS LSTM, an energy-efficient FPGA implementation for the inferencing phase of sparse LSTM networks, was proposed. Its architecture is compatible with the suggested pruning algorithm, which utilized two configurable processing elements with different sparsity ratios. It takes advantage of the sensitivity of the two different weight matrices to the pruning. Finally, the efficiency of the proposed pruning algorithm and accelerator was evaluated using selected benchmarks in NLP, sentiment classification, and speech recognition fields. Compared the state-of-the-art work, the proposed architecture and pruning algorithm provided, on average, 128% improvements in effective GOPS/W, and a 0.7% reduction in perplexity.

## REFERENCES

[1] A. Graves, A.R. Mohamed, and G. Hinton, "Speech Recognition with Deep Recurrent Neural Networks," *in Proc. of the IEEE Int. Conf. Acoustic Speech Signal Process.*, pp. 6645-6649, May 2013.

[2] H. Palangi et al., ''Deep sentence Embedding using Long Short-Term Memory Networks: Analysis and Application to Information Retrieval,'' *IEEE/ACM Trans. Audio, Speech, Language Process.*, vol. 24, no. 4, pp. 694–707, Apr. 2016.

[3] S. Hochreiter and J. Schmidhuber. "Long Short-Term Memory," Neural computation, 9(8):1735–1780, 1997.

[4] S. Han et al., "ESE: Efficient Speech Recognition Engine with Sparse LSTM on FPGA," Dec. 2016.

[5] A. Chang and E. Culurciello, "Hardware Accelerators for Recurrent Neural Networks on FPGA," in *Proc. of the IEEE Int. Sym. on Circuits and Systems (ISCAS)*, pp. 1-4, May 2017.

[6] Y. Guan et al., "FPGA-based Accelerator for Long Short-Term Memory Recurrent Neural Networks," in *Proc. of Asia and South Pacific Design Automation Conference*, pp. 629-634, 2017.

[7] E. Bank-Tavakoli, S.A. Ghasemzadeh, M. Kamal, A. Afzali-Kusha, and M. Pedram, "POLAR: A Pipelined/Overlapped FPGA-Based LSTM Accelerator," in *IEEE Transactions on Very Large-Scale Integration (VLSI) Systems*, 2019.

[8] S. Narang, E. Undersander, and G. F. Diamos, "Block-sparse Recurrent Neural Networks," in *CoRR*, 2017.

[9] S. Cao et al., "Efficient and Effective Sparse LSTM on FPGA with Bank-Balanced Sparsity," in *Proc. Int. Symp. Field-Programmable Gate Arrays*, pp. 63-72, 2019.

[10] S. Han, H. Mao, and W. J. Dally, "Deep Compression: Compressing Deep Neural Networks with Pruning Trained Quantization and Huffman Coding," in *Proc. ICLR*, 2016.

[11] S. Han et al., "Learning Both Weights and Connections for Efficient Neural Networks," in *Proc. NIPS*, pp. 1135-1143, 2015.

[12] W. Wen et al., "Learning Structured Sparsity in Deep Neural Networks," in *Proc. NIPS*, pp. 2074-2082, 2016.

[13] H. Mao et al., "Exploring the Regularity of Sparse Structure in Convolutional Neural Networks," in *Proc. CVPR Workshop Tensor Methods in Comput. Vis.*, 2017.

[14] S. Wang et al., "C-LSTM: Enabling Efficient LSTM Using Structured Compression Techniques on FPGAs" in *Proc. of the 2018 ACM/SIGDA Inter. Symp. on Field-Programmable Gate Arrays, ACM*, pp. 11-20, 2018.

[15] V. Y. Pan, "Structured Matrices and Polynomials: Unified Superfast Algorithms," *New York: Springer*, 2001.

[16] C. Gao et al., "DeltaRNN: A Power-efficient Recurrent Neural Network Accelerator," in *Proc. of the 2018 ACM/SIGDA Inter. Symp. on Field-Programmable Gate Arrays, ACM*, 21–30, 2018.

[17] M. Marcus et al. 1999. Treebank-3 LDC99T42. CD-ROM. Philadelphia, Penn.: Linguistic Data Consortium (1999).

[18] IMDb Datasets. https://www.imdb.com/interfaces

[19] J. S. Garofolo et al. Darpa TIMIT Acoustic-Phonetic Continous Speech Corpus CD-ROM. NIST Speech Disc 1-1.1. NASA STI/Recon technical report N, 93.

[20] J. Park et al., "Maximizing System Performance by Balancing Computation Loads in LSTM Accelerators," *Design Aut. Test in Europe Conf. Ex. (DATE)*, pp. 7-12, March 2018.

[21] Chip Huyen, "Evaluation Metrics for Language Modeling," The Gradient, 2019.

[22] S. Han et al., "EIE: Efficient Inference Engine on Compressed Deep Neural Network," in *Proc. ISCA*, pp. 243-254, 2016.